\documentclass[10pt,journal,compsoc]{IEEEtran}
\DeclareUnicodeCharacter{2212}{-}

\usepackage{ifpdf}
\usepackage{comment}

\usepackage{enumitem}
\usepackage{varwidth}
\usepackage{amsmath}

\usepackage[table]{xcolor}

\usepackage{tikz}

\usepackage[framemethod=tikz]{mdframed} %

\ifCLASSOPTIONcompsoc
\usepackage[nocompress]{cite}
\usepackage{comment}

\else
\usepackage{cite}
\fi

\usepackage{soul}

\newif\ifrevising
\revisingtrue 

\makeatletter
\newcommand*{\rom}[1]{\expandafter\@slowromancap\romannumeral #1@}
\makeatother

\begin{document}
%
\title{Implicit Mentoring: The Unacknowledged Developer Efforts in Open Source}
%
%
%
%

\author{Zixuan Feng,
        Amreeta Chatterjee,
        Anita Sarma,
        and~Iftekhar Ahmed
\thanks{Zixuan.Feng is with Electrical Engineering and Computer Science of
Oregon State University (e-mail: fengzi@oregonstate.edu).}
\thanks{Amreeta Chatterjee is with Electrical Engineering and Computer Science of
Oregon State University (e-mail: chattera@oregonstate.edu).}
\thanks{Anita Sarma is with Electrical Engineering and Computer Science of
Oregon State University (e-mail: anita.sarma@oregonstate.edu).}
\thanks{Iftekhar Ahmed is with Information and Computer Science at the University of California, Irvine.  (e-mail: iftekha@uci.edu).}}
\IEEEtitleabstractindextext{%
\begin{abstract}
Mentoring is traditionally viewed as a dyadic, top-down apprenticeship. This perspective, however, overlooks other forms of informal mentoring taking place in everyday activities in which developers invest time and effort but still need to be acknowledged. Here, we investigate the different flavors of mentoring in Open Source Software (OSS). In our previous work, we defined implicit mentoring---situations where contributors guide others through instructions and suggestions embedded in every day (OSS) activities. Through an empirical investigation of Pull Requests (PRs) in 37 Apache Projects, we built a classifier and extracted implicit mentoring comments. In this work, we continue to investigate implicit mentoring by analyzing it through the dual perspectives of experience and gender. Our analysis of 107,895 PRs shows that implicit mentoring occurs (27.41\% of all PRs include implicit mentoring) and does not follow the traditional dyadic, top-down apprenticeship model. When considering the gender of mentor-mentee pairs, we found pervasive homophily--a preference to mentor those of the same gender. In cross-gender mentoring instances, women were more likely to mentor men.
\end{abstract}

}

\maketitle

\IEEEdisplaynontitleabstractindextext

%
\IEEEpeerreviewmaketitle

\section{Introduction}
\label{sec:intro}

\IEEEPARstart{O}SS projects often adopt mentorship to assist contributors overcome barriers to contribution \cite{balali2020recommending, steinmacher2013newcomers, balali2018newcomers, sarma2016training}. Mentoring is an interpersonal connection in which an experienced individual gives an inexperienced individual with functional assistance and social guidance~\cite{kram1988mentoring}. These mentoring relationships plays a crucial role in ensuring the long-term viability of OSS projects by training new (and current) contributors who need to learn the required technical skills, the community's procedures, and cultural norms \cite{silva2020google}. Research shows that mentoring is an effective means for newcomer training and improves the on-boarding experience and retention of contributors in OSS \cite{fagerholm2014role,schilling2012train}.

Large OSS foundations have heavily invested time and efforts in formal mentorship programs for onboarding newcomers (e.g., the Apache Mentorship Program) \cite{apachementoring}. A total of 19,000+ new OSS contributors from 112+ countries have been paired with 18,000+ mentors from 133+ countries through the Google Summer of Code program since its inception in 2005 \cite{googlecode}. Linux, as one of the other global leaders in OSS, has launched seven mentorship programs and invested over \$2.5 million to support new OSS contributors \cite{linux}.

Outside of such formal mentoring, contributors actively seek guidance and support from each other via informal channels such as direct contact through emails and video conferencing or meeting at conferences. Moreover, technical guidance is also provided in everyday development activities, such as when contributors review code or design.

Irrespective of the type of mentoring, there is a cost involved in mentoring. Mentors have to spend time and effort in guiding newcomers and can face various challenges. For example, recommending a task that is suited to newcomers' background, interests, and matching the project timeline can be difficult~\cite{balali2020recommending}. It can also be challenging for mentors to keep the mentee engaged, especially if the project culture is harsh or the mentees are not proactive/ strongly interested in OSS~\cite{balali2018newcomers}. Given that mentors are also volunteers, this additional effort in mentoring can lead to a reduction in their technical productivity~\cite{fagerholm2014onboarding} and sometimes even loss in stature, where women who frequently mentor may be treated as community managers and not engineers~\cite{balali2018newcomers}. Currently, not much is known about the different types of mentoring in OSS. Without such an understanding, the important contributions made by OSS mentors may go unacknowledged.

One way to overcome this issue is to acknowledge (and promote) informal mentoring that can implicitly occur in everyday technical activities, such as code reviews. It is well known that code reviews are not always supportive. Past work has found that the ratio of negative sentiments is higher than that of positive sentiments in code review comments~ \cite{paul2019expressions}, and that destructive criticism, negative feedback that is nonspecific and inconsiderate, is fairly common~\cite{gunawardena2022destructive}. Thus, acknowledging those who take the additional effort to mentor by providing constructive feedback and explanation when suggesting changes or improvement is worthwhile. Such recognition is important not only to encourage ``implicit'' mentors, but also to sustain this mode of mentoring, which potentially: 
(1) requires less effort than that needed for a dedicated mentor-mentee relationship, 
(2) is topical and aligned with the mentor's technical interests,
(3) is part of the mentor's development activity, and 
(4) allows both the mentors and mentees to ``learn on the job''.

In our recent work \cite{feng2022}, we sought to understand the following aspects of implicit mentoring in OSS projects. We first defined implicit mentoring by performing a review of related work on mentoring, and then through formative interviews with OSS contributors and member-checking. We then empirically investigated the occurrence and impact of implicit mentoring via Pull Requests (PRs) in $37$ Apache projects. Finally, we surveyed 231 developers who had contributed to these projects to triangulate our results. This research was guided by the following research questions:

\noindent\textbf{RQ1: How can mentoring be implicitly provided in everyday development tasks?}

\noindent\textbf{RQ2: How prevalent is ``implicit mentoring'' in OSS projects?}

\noindent\textbf{RQ3: How does implicit mentoring impact OSS contributors---mentors and mentees?}

However, we still do not have a clear understanding of the extent to which implicit mentoring occurs and under what conditions? Who provides implicit mentoring? Whether it follows the traditional dyadic models of experienced mentors guiding newcomers? Whether there are any gender imbalances in who is mentoring or who gets mentored? Answers to these questions can help OSS projects better recognize implicit mentors and address any imbalances. Therefore, this journal extension continues the above investigation by exploring the following research questions:

\noindent\textit{\textbf{RQ4: What are the characteristics of implicit mentoring relationships?}}

\noindent\textit{\textbf{RQ5: What role do women play in implicit mentoring?}}


Our results lay the foundation for various future research directions and also identifies several calls for OSS communities such as creating an appreciative community, improving the state of diversity in OSS, and mechanisms to make mentoring sustainable; both of which are important to sustain and create healthy OSS projects. In the words of one of our interviewees : ``\textit{in open source, you're gonna find a lot of professionals that have a lot of experience, but need mentorship to understand and join a specific technology...but I don't think the mentorship is recognized at all.} [P3\footnote{P(N) refers to interview participant number.}]''

\section{Related Work}
\label{sec:relatedwork}

Mentoring has been extensively researched in various domains. In management and organizational literature, works have investigated the extent to which mentoring helps with organizational citizenship behaviors (defined as positive employee attitude to the organization)~\cite{eby2015cross}. They found top-down mentoring to be positively associated with improved employee attitudes. Payne and Huffman~\cite{payne2005longitudinal} investigated the relationship between mentoring and positive organizational attitudes and found it to have a strong association with effective commitment (employee's emotional attachment or identification with the organization) and continuance commitment with the organization. Similarly, in education literature, Mullen and Klimaitis~\cite{mullen2021defining} conducted a literature review of empirical studies on mentoring to identify other forms in addition to the traditional---formal, dyadic, top-down---mentoring model. These included student mentoring that is done in groups, among peers, in collaborative or cross-cultural forms.

Another form of mentoring is informal mentoring. According to Inzer and Crawford~\cite{inzer2005review}, informal mentoring occurs in a relationship between two people where one gains insight, knowledge, wisdom, friendship, and support from the other. Ko et al.~\cite{ko2018informal} found informal mentoring to be better at triggering and maintaining interest in computing among adolescent students. Nandi and Mandernach~\cite{nandi2016hackathons} concluded that students are strongly motivated by informal mentoring relationships. Analysis from surveys, student academic records and source-code commit log data show that students improve significantly when informally mentored.

Multiple works have found informal mentoring to be more beneficial than formal mentoring as it provides higher engagement and skill development opportunities, including coaching, providing challenging assignments, or increasing exposure and visibility of the mentee. Mentees who were informally mentored were much more satisfied with their mentors than mentees with formal mentors~\cite{baugh2007formal, inzer2005review,bynum2015power}. Ragins and Cotton~\cite{ragins1999mentor} found women had the least to gain from formal mentoring, as the presence of formal mentors reduced coaching, role modelling, and career counselling for women mentees. Informal mentoring on the other hand was beneficial for both men and women. Inzer and Crawford~\cite{inzer2005review} also state that informal mentoring is a valuable tool for grooming an employee as it occurs in a relationship that is voluntary and created by both persons.

In our recent work \cite{feng2022}, we defined a type of informal mentoring in the context of OSS, i.e., implicit mentoring, as well as analyzed its prevalence and impact. Results showed that implicit mentoring is prevalent in daily development activities in OSS, and both mentees and mentors benefit from it. However, there still remains a gap in our understanding of the characteristics of implicit mentoring, which we aim to fill by investigating how implicit mentoring happens and who these mentors are in the OSS community.


\section{method}

\begin{figure*}[!tbp]
\centering
\includegraphics[width=7.1in]{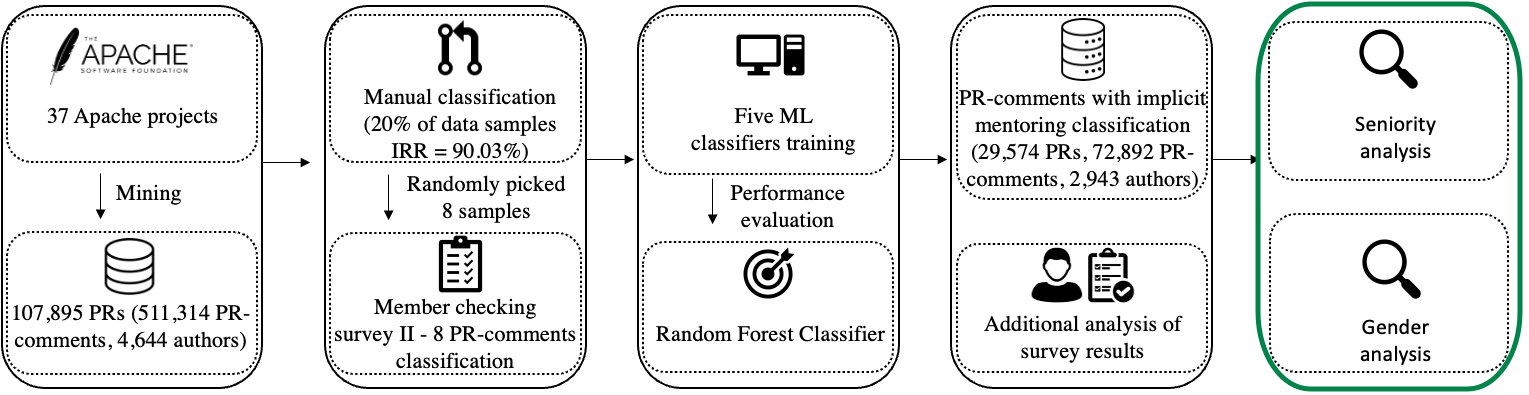}
\caption{Overview of research method}
\label{fig:method}
\end{figure*}

Figure \ref{fig:method} shows an overview of our methodology, with the highlighted ``green" box representing the methods followed in this extension. Prior to this (as shown in the non-highlighted portion), we conducted a literature review on mentoring, followed by formative interviews with OSS contributors, member-checking surveys, and large-scale surveys with ASF contributors to define implicit mentoring~\cite{feng2022}. We then created a classifier to extract implicit mentoring PR comments from 37 Apache Projects. In this section, we will briefly discuss our previous work, and explain the methods we used to answer the two RQs addressed in this extension.

Table \ref{tab:apache} presents the statistics of the 37 projects in our dataset. The dataset includes projects with large codebases (Table \ref{tab:apache} specifies project size in KLOC) and contributors. This dataset had 107,990 PRs and 836,729 PR-comments from 12,668 authors. Our research requires GitHub profile data, so we excluded contributions from 42 user profiles whose accounts were deactivated during data collection. In addition, we filtered out PR-author comments because we only intended to assess mentorship comments from other contributors. This led to 107,895 PRs, 511,314 PR-comments, and 12,626 contributors (out of which 4,644 were PR-comment authors).

\begin{table}[tb]
\caption{Project statistics of the 37 Apache Projects.}
\resizebox{\columnwidth}{!}{
\begin{tabular}{lllll}
\textbf{Dimension}               & \textbf{Max} & \textbf{Min} & \textbf{Average} & \textbf{Median} \\ \hline
\rowcolor[HTML]{EFEFEF} 
\textbf{Project size (KLOC)}              & 18,475   & 82       & 1,770        & 1,075       \\
\textbf{Project Age (weeks)}                   & 1,063        & 214          & 606              & 560             \\
\rowcolor[HTML]{EFEFEF} 
\textbf{Developers}              & 1,852        & 21           & 226              & 106             \\
\textbf{Total Commits}           & 80,227       & 3,561        & 22,688            & 18,084          \\
\rowcolor[HTML]{EFEFEF} 
\textbf{Total Pull-request (PR)}  & 32,645       & 6            & 2916            & 651             \\
\textbf{Total PR-comments(Non-PR author)}        & 316,295      & 2            & 14609            & 807           
\end{tabular}}
\label{tab:apache}
\end{table}

\textit{\textbf{Manual Classification of Sample Data}}: 
To answer our research questions, we needed to differentiate between PR comments with and without implicit mentoring. We employed machine learning to classify 511,314 PR comments. To train the machine learning classifier, we manually classified a training set. We used a 95\% confidence interval and a 5\% margin of error to calculate the size of this training dataset \cite{kotrlik2001organizational}, which was 384 PR comments. Then, we randomly chose 384 PR comments. The first two authors manually labeled a subset of PR comments in the training set to calculate Inter-Rater Reliability (IRR) \cite{hallgren2012computing}. A PR comment was considered to contain implicit mentoring if it included an ``explanation''  in addition to giving suggestions, instructions, or helping fix errors (see ~\ref{results}). Table~\ref{tab:rulebook} shows the rule book we used with examples of PR comments. The two authors independently labeled 20\% of the PR-comments and reached high IRR (90.03\% Cohen Kappa~\cite{viera2005understanding}). The remaining 80\% of the training dataset was split evenly between the two authors who manually classified the PR-comments.

\begin{table}[tbp]
\caption{Classification rule book.}
\resizebox{\columnwidth}{!}{
\begin{tabular}{l|l|l}
\textbf{Mentoring} & \multicolumn{1}{c|}{\textbf{PR-comment sample}}                                                                                                                                                                      & \multicolumn{1}{c}{\textbf{\begin{tabular}[c]{@{}c@{}}Mentoring \\ Action\end{tabular}}} \\ \hline
\rowcolor[HTML]{EFEFEF} 
YES                & \begin{tabular}[c]{@{}l@{}}“...run {[}tool{]} on the project before\\ creating a PR. You would have noticed {[}problem{]}...”\end{tabular}                                                                     & Instruction                                                                              \\
YES                & \begin{tabular}[c]{@{}l@{}}“ I would still duplicate {[}action{]} like I did\\ in {[}certain PR{]} because it’s widely used in {[}tests{]}.\\ Maybe this could be removed after {[}situation{]}.”\end{tabular} & Suggestion                                                                               \\
\rowcolor[HTML]{EFEFEF} 
YES                & \begin{tabular}[c]{@{}l@{}}“ Would you mind just doing {[}action{]}\\ again to kick off {[}framework{]}? I think {[}framework{]}\\ is just not happy when it has a lot of loads.”\end{tabular}                & \begin{tabular}[c]{@{}l@{}}Mechanisms\\ to fix errors\end{tabular}                       \\
NO                 &"LGTM, Merging into master and release-[version number]"                                                                                                                                                                 & NA                                                                                      \\
\rowcolor[HTML]{EFEFEF} 
NO                 & \begin{tabular}[c]{@{}l@{}}"This PR currently has merge conflicts, but {[}\#PR{]} \\ is next in line, so you may want to wait till it is merged \\ before you fix these conflicts."\end{tabular}              & NA                                                                                      \\
NO                 & \begin{tabular}[c]{@{}l@{}}"This is not ready. Missing apache header \\ on the new file and there is no test. No idea what \\ this is fixing."\end{tabular}                                                    & NA                                                                                     
\end{tabular}}
\label{tab:rulebook}
\vspace{-4mm}
\end{table}

\textit{\textbf{Member Checking Survey:}} To ensure the validity of our manual classification, we contacted the five interviewees (Table \ref{tab:participant_demo} presents the demographic information of our participants) to validate our PR-comment labeling. We randomly selected eight PR-comments (four implicit mentoring, four not implicit mentoring) from the manually classified set of 384 PR-comments. The survey was created in Qualtrics~\cite{2015Qualtrics} and conducted face-to-face using Zoom~\cite{archibald2019using}. The survey presented the PR-comment and included three response options: Mentoring, Not mentoring, and Not sure. For the first two options, participants were asked to give a reason for their response. Participants could either type in or verbalize their answers (See supplementary~\cite{supply} for further details about Member Checking Survey).

\begin{table}[!tbp]
\caption{Demographic information of interview participants.}
\resizebox{\columnwidth}{!}{
\begin{tabular}{ccc|cccc}
\multicolumn{1}{l}{} & \multicolumn{1}{l}{} & \multicolumn{1}{l|}{}    & \multicolumn{4}{c}{\textbf{Mentoring experience}}                                                                                                                                                  \\ \hline
\textbf{ID}          & \textbf{Gender}      & \textbf{OSS experiences} & \textbf{Mentor} & \multicolumn{1}{c|}{\textbf{\begin{tabular}[c]{@{}c@{}}Informal\\ /Formal\end{tabular}}} & \textbf{Mentee} & \textbf{\begin{tabular}[c]{@{}c@{}}Informal\\ /Formal\end{tabular}} \\
\rowcolor[HTML]{EFEFEF} 
P1                   & Woman                & Over 10 years            & Y               & \multicolumn{1}{c|}{\cellcolor[HTML]{EFEFEF}Both}                                        & Y               & Both                                                                \\
P2                   & Man                  & Over 10 years            & Y               & \multicolumn{1}{c|}{Both}                                                                & Y               & Both                                                                \\
\rowcolor[HTML]{EFEFEF} 
P3                   & Woman                & 6-10 years               & Y               & \multicolumn{1}{c|}{\cellcolor[HTML]{EFEFEF}Both}                                        & Y               & Both                                                                \\
P4                   & Woman                & 6-10 years               & Y               & \multicolumn{1}{c|}{Both}                                                                & Y               & Both                                                                \\
\rowcolor[HTML]{EFEFEF} 
P5                   & Woman                & 6-10 years               & Y               & \multicolumn{1}{c|}{\cellcolor[HTML]{EFEFEF}Both}                                        & Y               & Both                                                               
\end{tabular}}
\label{tab:participant_demo}
\vspace{-4mm}
\end{table}

The member checking survey responses were 90\% in agreement with our classification (the remaining 10\% was when participants selected ``Not Sure''). This suggests that our manual classification is reliable. Moreover, their explanations in the open-ended replies indicate that our rule book is reliable. For example, P3 explained why a PR-comment was mentoring: ``\textit{This is helpful feedback in explaining that these changes are of low value}'' And P4 explained why a PR-comment was not mentoring: ``\textit{This is asking for information [and not mentoring]}''.

\textit{\textbf{Machine Learning Classifier}}: Using the manually classified corpus, we trained five common supervised  machine learning classifiers as they have been used in similar studies for classifying discussion comments~\cite{mannan2020relationship, brunet2014developers}. They were: Random Forest ~\cite{pal2005random}, Bernoulli~\cite{singh2019comparison}, Support Vector Machine~\cite{gunn1998support}, KNeighbors~\cite{peterson2009k}, and Decision tree~\cite{kothari2001decision}. Our training data consisted of the comments in the PR. We used the usual text cleaning steps which included using Porter's stemming~\cite{porter1980algorithm}. We eliminated all of the terms from the standard stop word list in order to do suffix stripping. To ensure the best performance, we applied hyper-parameter adjustments from the Python Scikit learn library~\cite{pedregosa2011scikit} to all five classifiers. By applying $ScikitLearnRandomizedSearchCV$, we found the optimal parameters for each classifier. The models were trained and evaluated using a 10-fold cross validation methodology. That is, the data was randomly divided into 10 equal sets, and nine sets were used for training and one for testing. We trained our model using this method 10 times and report the mean scores.

\begin{table}[tbp]
\caption{Classification results per classifier.}
\resizebox{\columnwidth}{!}{
\begin{tabular}{lllll}
                                & \textbf{Precision} & \textbf{Recall} & \textbf{F1} & \textbf{AUC} \\ \hline
\rowcolor[HTML]{EFEFEF} 
\textbf{RandomForest}           & 0.87               & 0.90            & 0.88        & 0.94         \\ \hline
\textbf{Support Vector Machine} & 0.84               & 0.84            & 0.84        & 0.91         \\ \hline
\rowcolor[HTML]{EFEFEF} 
\textbf{NaiveBayes}             & 0.81               & 0.79            & 0.78        & 0.90         \\ \hline
\textbf{DecisionTree}           & 0.74               & 0.74            & 0.73        & 0.81         \\ \hline
\rowcolor[HTML]{EFEFEF} 
\textbf{K-neighbors}            & 0.72               & 0.62            & 0.57        & 0.71        
\end{tabular}}
\label{tab:ml}
\vspace{-4mm}
\end{table}

Table~\ref{tab:ml} shows the precision, recall, F1, and AUC scores of the five classifiers~\cite{powers2011evaluation}. Random Forest Classifier (RFC) had the best overall performance when considering both the F-measure (0.88) as well as the AUC scores (0.94). Therefore, we used RFC for further analysis. 
The final tuned parameters for our classifier RFC were $n\_estimators$=$2800$, $max\_features$=$auto$, $max\_depth$=$73$, $min\\\_samples\_split$=$20$,$min\_samples\_leaf$=$2$, and $bootstrap$=$True$.

\begin{figure}[!tbp]
\centering
\includegraphics[width=0.5\linewidth]{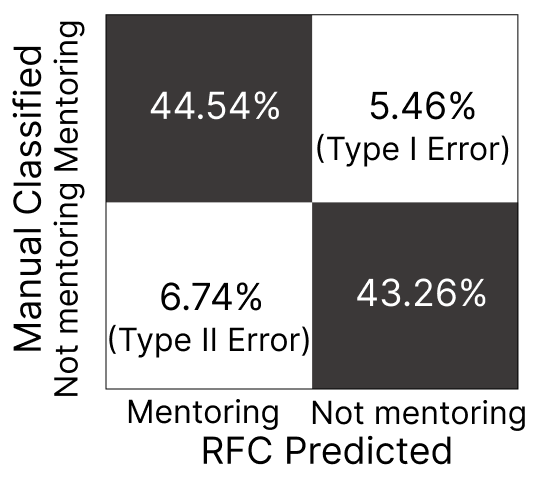}
\caption{Confusion Matrix.}
\label{fig:confusion_matrix}
\end{figure}

\textit{\textbf{Manual Analysis of Misclassification:}} Any systemic error from misclassified PR-comments could be critical to its acceptance. We investigated the confusion matrix~\cite{visa2011confusion} of our RFC model as shown in Figure \ref{fig:confusion_matrix}. We can infer from the confusion matrix that neither class (Mentoring, Not-mentoring) is disproportionately impacted by the RFC's misclassifications.

\textit{\textbf{Large scale online survey:}} To acquire a more generalized perception about implicit mentoring, we surveyed OSS developers to gain a broader understanding of implicit mentorship. We surveyed Apache OSS developers since Apache Software Foundation is dedicated to promoting and enhancing mentorship in OSS (Google Summer of Code~\cite{silva2020google}). Additionally, Apache has explicit guidelines for mentors that contributors have to adhere to~\cite{apachementoring}, making Apache developers a suitable target for our survey.

\textit{Survey design:} Our survey comprised 19 questions, a mix of multiple-choice, Likert scale, and open-ended questions (see supplementary for the survey questions~\cite{supply}). The survey included demographics questions (Q1-Q5), validation questions of implicit mentoring definitions (Q6, Q9), participant experiences/satisfactions with it (Q7-Q8, Q13-Q14, Q18-Q19), and finally questions about participants' perceptions of the impact of implicit mentoring (Q10-Q12, Q15-Q17). We conducted five pilot studies with graduate students and professionals with OSS experience using snowball sampling~\cite{goodman1961snowball}. After each pilot study, we collected feedback and refined the survey.

\textit{Survey responses:} We used the GitHub API~\cite{mombach2018github} to mine contributor emails from these 37 projects. After removing emails of accounts that were either deleted or private, we were left with 3,699 developer email addresses in total. We used Qualtrics~\cite{2015Qualtrics} as a distribution platform to deploy our survey. We emailed the survey to these 3,699 developers (following university-approved IRB protocol), and 70 emails bounced (giving 3,629 valid emails). The survey was open for two weeks, during which we received 231 responses or a response rate of 6.37\%. These response rates are consistent with other studies in software engineering~\cite{wessel2020expect}.

\textit{\textbf{Identifying Experiences of Mentors and Mentees}}: 
To investigate the extent to which top-down interactions comprise implicit mentoring (RQ4), we needed to first identify whether someone is a mentee or a mentor and how experienced they are. We considered the PR author as a mentee if any of the PR-comments associated with that PR was classifed as ``implicit mentoring''. For those PR-comments identified as implicit mentoring, their comment authors were labelled as mentors.

To investigate the dynamics of implicit mentoring, especially if it comprises top-down interactions (experienced to inexperienced) we analyzed the characteristics of implicit mentoring at three levels based on the time and amount of contributions made. Our measurements of amount and time are based on Quinones et al.'s conceptualization of work experience~\cite{quinones1995relationship}. 

\textbf{(a)} experience within the specific project, which we calculated based on the difference between the dates of the mentor and mentee's first contribution (PR or PR-comment) within the project. 
\textbf{(b)} However, in OSS, contributors can participate in multiple projects or can leave a project to join another. In such cases, they may bring experiences from working on another project when mentoring a developer in a project. Therefore, we also calculated the overall experience of a contributor in GitHub based on the date on which they created their GitHub account.
\textbf{(c)} In addition, we also used the amount of contribution (PRs) as another proxy to investigate the dynamics of implicit mentoring. We calculate the amount of PRs submitted before the current PRs for both mentees and mentors within the project.

We then analyzed survey responses from ASF contributors to validate our analysis. In the survey, we had asked respondents how long they had participated in OSS (Q5) and how long they had been implicit mentors/mentees in OSS (Q7). 

We compared these responses (how long they had been a mentee/mentor vs. how long they had been participating in OSS) to validate whether their implicit mentoring experience followed a traditional top-down hierarchy, meaning they were implicit mentees for a lesser time than they were implicit mentors/OSS experiences.

\textit{\textbf{Identifying Gender}}: 
To investigate the gender diversity in implicit mentoring (RQ5), we used the ``Namsor" API to identify the gender of contributors in our dataset. ``Namsor" is a name recognition API that estimates the gender of a (full) name on a -1 to +1 probability scale based on geographic information~\cite{namsor}. Multiple studies have addressed the reliability of the ``Namsor" gender classifications, the error is less than 10\%\cite{santamaria2018comparison, santamaria2018comparison}.

However, using GitHub profile-reported location for gender prediction results in inaccuracy. For instance, the accuracy of gender prediction based on an Asian name is negatively affected if the region is reported as North America. Moreover, not all contributor profiles provide precise location information. Therefore, our approach involves utilizing ``Namsor" to first predict the original region associated with the name, followed by using both the predicted region and name to make the gender prediction. Our final gender-analysis dataset consists of 6,224 contributors (6,027 men contributors, 197 women contributors), for whom ``Namsor" gender identification probability was over 90\%.

\section{Results}
\label{results}
\subsection{Definition of Implicit mentoring}

Recall, we defined implicit mentoring as mentoring that occurs during OSS routine development activities, such as code reviews, in which a mentor provides an underlying explanation while providing recommendations, directions, or mechanisms to address mistakes. This kind of mentoring can occur through a variety of channels, including emails, PR-comments, in-person meetings, and online communication tools \cite{feng2022}.

To validate our definition of implicit mentoring, we surveyed 231 ASF contributors. There were five possible responses, ranging from "Strongly agree" to "Strongly disagree." We considered a respondent to be in agreement with the definition if they selected "Agree" or "Strongly agree." The majority (214 out of 231, 92.64\%) of respondents agreed with our definition of implicit mentoring in OSS. 

\subsection{Identifying implicit mentoring in OSS projects:}

We used the RFC classifier on our data set to identify the frequency of implicit mentoring in PR comments. Identifying such mentors has two direct benefits. First, an ability to (formally) acknowledge the effort of mentors, as P3 stated: ``\textit{people within [project], who have dedicated their entire careers to mentor interns and they don't get recognized for it.}'' Second, projects currently struggle to identify mentors, as P5 said: ``\textit{our leads are burned out by too much of [mentoring]...I'm trying to figure out how can I identify the people who've been mentored or who have been at the intermediate level, and get them engaged in the mentoring.}''

In our dataset, a PR received 4.74 PR-comments on average ($sd=8.81$), which were not made by the PR-author. It is possible that one or more of these PR-comments could embed implicit mentoring. Overall, 27.41\% of PRs included implicit mentoring comments. These comments were made by 2,943 out of 4,644 PR-comment authors (63.37\%).

\subsection{Interaction Types in Implicit Mentoring}
\label{flavors}

\begin{figure}[!tbp]
\centering
\includegraphics[width=\columnwidth]{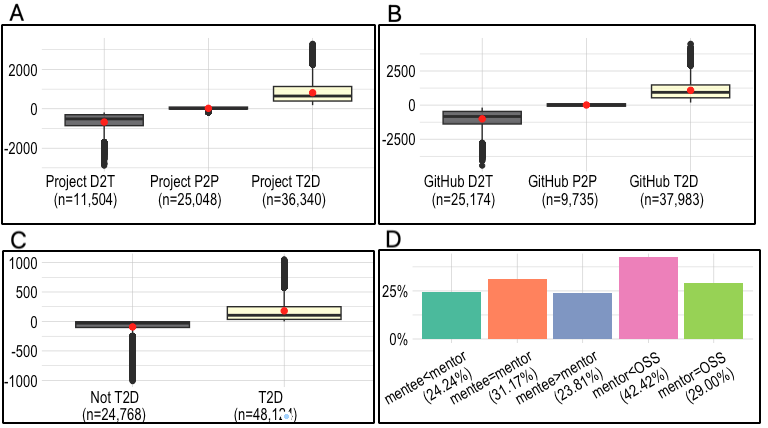}
\caption{(A) Difference in experiences of mentor/mentee (Project); (B) Difference in experienes of mentor/mentee (GitHub); (C) Difference in amount of contributions of mentor/mentee; (D) OSS experiences vs. Mentoring experience (Survey)}
\label{fig:PR}
\end{figure}

Traditionally, mentoring has been viewed as a dyadic relationship wherein an experienced individual (the mentor) provides practical advice and guidance to an inexperienced individual (the mentee)~\cite{Kram_1988, gandhi2012reverse}, helping them gain technical and job-related skills~\cite{brashear2006exploratory, hale2000match}. Therefore, we wanted to analyze if the same sort of mentoring dynamics occur in implicit mentoring, leading to:\\
\textit{\textbf{RQ $\#4$:} To what extent is implicit mentoring characterized by dyadic, top-down interactions?}

We answer the first part of the question by considering the interactions for each PR. The PR-author is considered the mentee. Contributors of PR-comments that included implicit mentoring are considered mentors. If the discussions occurred only between the mentee and one mentor, this implicit mentoring was considered \textit{dyadic}. Similarly, if the discussions occured between a mentee and two mentors, its considered a triad. Finally, we group all mentee-mentor groups equal to or greater than four as $\ge$quadrad. Note, that for a PR, if there were other contributors who commented, but were not mentors (PR-comment was not considered implicit mentoring), they are ignored in this analysis.

Of the 29,574 PRs with implicit mentoring, the majority was dyadic (66.79\%), lining up with the traditional view of mentor-mentee relationship. However, because of the open, voluntary-based nature of OSS, we found 21.85\% of implicit mentoring occurring in triads and another 11.36\% in $\ge$quadrads. This is akin to ``it takes a village...'' adage, where the community of contributors work together to mentor new contributors and providing perhaps different perspectives and guidance.

\vspace{0.3cm}
\begin{mdframed}[roundcorner=10pt]
\textbf{Observation 1:} Implicit mentoring does not follow to the conventional dyadic connection, and often involves multiple mentors supporting a single mentee.
\end{mdframed}
\vspace{0.3cm}

Next, we investigate what role experience---within the project and overall---plays in implicit mentoring. There can be three types of mentoring directionality:

\begin{itemize}
    \item \textit{top-down}, the mentor is more experienced than the mentee,
    \item \textit{peer-to-peer}, the mentor and mentee have similar level of experience,
    \item \textit{bottom-up}, mentee is more experienced than the mentor.
\end{itemize}

To obtain the approach of implicit mentoring, we first start by getting the first date of contribution (a PR or a PR-comment) for both mentors and mentees. Then, we subtracted the date of first contribution made by the mentee from that of their respective mentors.

We use 6-months (183 days) as a threshold to classify the mentoring approaches. If the difference is greater than 6-months, then the PR-comment interaction is flagged as \textit{top-down} and if less than 6-months, it is \textit{bottom-up}. If mentor and mentee had their PR/PR-comments date within 6-months, we classify it as \textit{peer-to-peer}. Figure \ref{fig:PR}A and B shows the distribution of implicit mentoring when considering within the project as well as overall experience, where we use the GitHub account creation date as a proxy.

About 49.85\% (36,340) of implicit mentoring falls in the traditional top-down mentoring interaction style. The average difference between mentee and mentor experience was 2.24 years (817 days, $sd=536$). 
34.36\% (25,048) implicit mentoring was in peer-to-peer category, the mean difference between mentor and mentee was 25 days ($sd=83$). 15.78\% (11,504) of implicit mentoring was in the bottom-up category with the mean difference between mentor-mentee being 1.84 years (673 days, $sd=488$).

In OSS, contributors frequently contribute to multiple projects or migrate across projects~\cite{jergensen2011onion}. In such cases, they may accrue skills and experiences relevant to the project elsewhere. Therefore, we look at the mentoring directionality when considering the overall experience in GitHub. 
The results (Figure \ref{fig:PR}B) show that the traditional top-down mentoring instances are 52.11\% (37,983) out of 72,892 PR-comments, the mean differences in experience being 2.96 years (1,082 days, $sd=682$). This was followed by \textit{peer-to-peer} mentoring (13.36\%) with the mean difference being 6 days ($sd=104$); and \textit{bottom-up} mentoring (34.54\% of cases) with mean difference in experience being 2.76 years (1,009 days, $sd=675$).

We then validated our findings by analyzing the experiences of the contributors with regards to the quantity of their contributions. This involved comparing the number of PRs made by implicit mentors and mentees. As shown in Figure \ref{fig:PR}C, in 66.02\% (48,124) of 72,892 mentoring occurrences, mentors had contributed more PRs than mentees before providing the current mentoring comments, with mean difference in contribution amount of 179.85 ($sd=199$). In 33.98\% of mentoring instances, mentees and mentors had contributed the same or greater contributions, with mean differences in amount of PRs of 92.68 ($sd=154$).

In contrast to conventional mentoring, implicit mentoring is applicable to OSS contributors from all experience levels, as also indicated by the survey results as shown in Figure \ref{fig:PR}D. Analysis from our survey indicated that 175 of 231 (75.76\%) respondents have been implicit mentors, and 67 (29.00\%) of them had have been mentors for as long as they had been in OSS. Therefore, unlike traditional mentoring, 29.00\% of respondents who begun to contribute to OSS also started to provide support to other contributors. This finding was in line with one of the comments from our survey respondents: \textit{``Being a mentor in OSS is all about sharing your experience, regardless the background. It is also tremendously self-rewarding as you gain popularity which helps career path evolution''} [ASF-97 \footnote{ASF(N) refers to large scale survey respondent number.}].

Out of the 155 (67.10\%) survey respondents who served as both mentor and mentee, over 50\% reported having spent an nearly same amount of time in both roles or having been a mentor for a longer period, as shown in the Figure \ref{fig:PR}D, which indicates that implicit mentoring is not confined to a specific level of experience; while contributors provided mentoring to others, they also received mentoring. As one respondent mentioned: \textit{``There are many forms of knowledge that came through one's experiences, some knowledge blossomed and became mainstay while others died out; its important to realise that sharing and receiving knowledge helps build culture and community''} [ASF-60].

As a result of investigating the characteristics of implicit mentoring concerning experiences across the four analysis as shown in Figure \ref{fig:PR} A, B, C, and D, the story remains consistent: implicit mentoring is for all contributors and is not limited to a specific level of experiences. This shows that “\textit{seniority is not necessary for implicit mentoring}” [P3].

\vspace{0.3cm}
\begin{mdframed}[roundcorner=10pt]
\textbf{Observation 2:} Implicit mentoring is long-term mentoring for all contributors and is not confined to a specific level of experiences.
\end{mdframed} 
\vspace{0.3cm}

Past work has found that women, especially in OSS, are more tuned to community building roles and end up being mentors more often~\cite{balali2018newcomers}. This often translates to women being seen as community managers and losing their ``engineer'' voice in decision making~\cite{bosu2019diversity}. Further, given the gender imbalance in OSS (past work having found women contributors to range around 10\%~\cite{bosu2019diversity}) and past research having shown that same-gender mentor-mentee dyadic relationships are more common~\cite{kogovvsek2019effects}, it can mean that fewer women have mentors. Both of these phenomena disadvantage women in OSS. Therefore, we wanted to investigate if implicit mentoring is also disproportionately done by women, leading to the research question:

\begin{table}[!tbp]
\caption{Summary of implicit mentoring by gender.}
\resizebox{\columnwidth}{!}{
\begin{tabular}{llll}
 & \textbf{women} & \textbf{men} & \textbf{Total} \\ \hline
\rowcolor[HTML]{EFEFEF} 
\textbf{PR-comment developers} & 42 (2.22\%) & 1,842 (97.78) & 1,884 \\
\textbf{Implicit mentors} & 29 (2.33\%) & 1,216 (97.67\%) & 1245 \\
\rowcolor[HTML]{EFEFEF} 
\textbf{PR-comments} & 420 (0.27\%) & 157,018 (99.73\%) & 157,438 \\
\textbf{\begin{tabular}[c]{@{}l@{}}PR-comments w/\\ Implicit mentoring\end{tabular}} & 130 (0.50) & 25,617 (99.50\%) & 25,747
\end{tabular}}
\label{tab:genderoverall}
\end{table}
\textbf{RQ $\#5$. What role do women play in implicit mentoring?}

To answer this question, we analyzed a reduced dataset comprising 6,224 PR/PR-comment contributors (out of which 1,884 contributors were PR-comment authors). Recall, we only kept data of individuals where the ``Namsor"~\cite{namsor} API predicted the gender with high confidence ($>90\%$).

Table \ref{tab:genderoverall} shows the distribution of the gender of contributors in this dataset and their mentoring activity. The gender distribution shows that 2.22\% (42) were women PR-comment authors and the 97.78\% (1,842) men PR-comment authors. A majority of these contributors---29 women (2.33\% of women) and 1,216 men (97.67\% of men)---served as implicit mentors.

Past research has shown that the conventional view of mentoring is that women tend to do more mentoring \cite{sosik2000role, trinkenreich2021please, balali2018newcomers}. 
In our dataset, we see that trend with slightly more women (69.05\%) serving as mentors as compared to men (66.02\%). To investigate if the differences between the proportions of implicit mentoring comments provided by men compared to women are statistically significant, we performed a two sample Z-test of proportions~\cite{ld1993biostatistics} (see Table ~\ref{tab:gender_overall_Z} (Overall column)).

The results for \textit{overall implicit mentoring} show that differences are significant ($estimate$=0.15, $P$ value$<$0.001), with women providing implicit mentoring 15\% times more than men. Moreover, as the estimated ratio of differences is medium, we calculated the effect size using Cohen's d \cite{cohen2013statistical}. An effect size of $d=0.35$ (Table \ref{tab:gender_overall_Z}, Overall Column) indicates the differences are small~\cite{cohen2013statistical}.

\begin{table}[!tbp]
\caption{Proportion test comparing implicit mentoring provided by men vs. women through PR-comments.}
\resizebox{\columnwidth}{!}{
\begin{tabular}{lllll}
\rowcolor[HTML]{FFFFFF} 
\textbf{Gender} & \textbf{Overall} & \textbf{Top$\rightarrow$Down} & \multicolumn{2}{l}{\cellcolor[HTML]{FFFFFF}\textbf{Non-Top$\rightarrow$Down}} \\ \hline
\multicolumn{5}{c}{\textbf{Calcuated based on first contribution date}} \\ \hline
\rowcolor[HTML]{EFEFEF} 
\textbf{Z-score} & \multicolumn{1}{l|}{\cellcolor[HTML]{EFEFEF}8.10} & 4.46 & \multicolumn{2}{l}{\cellcolor[HTML]{EFEFEF}6.70} \\
\textbf{P-value} & \multicolumn{1}{l|}{$<$ 0.01} & $<$0.01 & \multicolumn{2}{l}{$<$ 0.01} \\
\rowcolor[HTML]{EFEFEF} 
\textbf{\begin{tabular}[c]{@{}l@{}}Estimated\\ differences\end{tabular}} & \multicolumn{1}{l|}{\cellcolor[HTML]{EFEFEF}0.15} & 0.13 & \multicolumn{2}{l}{\cellcolor[HTML]{EFEFEF}0.15} \\
\textbf{Cohen d} & \multicolumn{1}{l|}{0.35} & 0.31 & \multicolumn{2}{l}{0.36} \\ \hline
\multicolumn{5}{c}{\textbf{Calcuated based on amount of contribution}} \\ \hline
\rowcolor[HTML]{EFEFEF} 
\textbf{Z-score} & \multicolumn{1}{l|}{\cellcolor[HTML]{EFEFEF}-} & 1.95 & \multicolumn{2}{l}{\cellcolor[HTML]{EFEFEF}11.05} \\
\textbf{P-value} & \multicolumn{1}{l|}{-} &  0.05 & \multicolumn{2}{l}{$<$ 0.01} \\
\rowcolor[HTML]{EFEFEF} 
\textbf{\begin{tabular}[c]{@{}l@{}}Estimated\\ differences\end{tabular}} & \multicolumn{1}{l|}{\cellcolor[HTML]{EFEFEF}-} & 0.07 & \multicolumn{2}{l}{\cellcolor[HTML]{EFEFEF}0.19} \\
\textbf{Cohen d} & \multicolumn{1}{l|}{-} & 0.15 & \multicolumn{2}{l}{0.52} \\
\rowcolor[HTML]{EFEFEF} 
\multicolumn{5}{l}{\cellcolor[HTML]{EFEFEF}$H_0: P_1-P_2 = 0$, $H_a: P_1-P_2 \neq 0$, adjusted $\alpha=0.017$} \\
\multicolumn{5}{l}{$P_1$:P(PR-comments by men: Mentoring/All)} \\
\rowcolor[HTML]{EFEFEF} 
\multicolumn{5}{l}{\cellcolor[HTML]{EFEFEF}$P_2$:P(PR-comments by women: Mentoring/All)}
\end{tabular}}
\label{tab:gender_overall_Z}
\end{table}

\vspace{0.3cm}
\begin{mdframed}[roundcorner=10pt]
\textbf{Observation 3:} Overall, more women than men provide implicit mentoring, and the difference is medium.
\end{mdframed}
\vspace{0.3cm}

Next, we investigate the different implicit mentoring approaches, such as top-down and non-top-down mentoring (peer-to-peer, bottom-up). We integrated peer-to-peer and bottom-up mentoring into non-top-down mentoring, as traditional mentoring is classified as top-to-down \cite{Kram_1988, gandhi2012reverse}. Table \ref{tab:gender_overall_Z}  presents the results of two sample Z-test of proportions and the effect sizes. Since we perform multiple-hypotheses testing, we applied Bonferroni correction to adjust $P$ values~\cite{napierala2012bonferroni}, which gives an adjusted $\alpha=0.017$.

For analysis based on differences in initial contribution date between mentors and mentees, there are no significant differences between implicit mentoring done by men and women for Top$\rightarrow$Down ($estimate$ =0.13, $P$ value $<$ 0.01, Cohen $d=0.31$). 
When considering non-traditional approaches (\textit{peer and Bottom$\rightarrow$Up implicit mentoring}), more women than men did implicit mentoring ($estimate$=0.15, $P$ value$<$0.001), while the differences are significant, the effect size is medium (Cohen $d=0.36$).

For analysis based on differences in amount of contribution between mentors and mentees, there is significant differences between implicit mentoring done by men and women for Top$\rightarrow$Down ($estimate$ =0.07, $P$ value$=$0.05). The non-traditional approach yields the same results, more women than men did implicit mentoring ($estimate$=0.19, $P$ value$<$0.001), and the effect size is medium (Cohen $d=0.52$).

\vspace{0.3cm}
\begin{mdframed}[roundcorner=10pt]
\textbf{Observation 4:} There are significant differences between implicit mentoring done by men and women, with women tending to provide more non-traditional implicit mentoring.
\end{mdframed}
\vspace{0.3cm}

\begin{figure}[!tbp]
\centering
\includegraphics[width=3.4in]{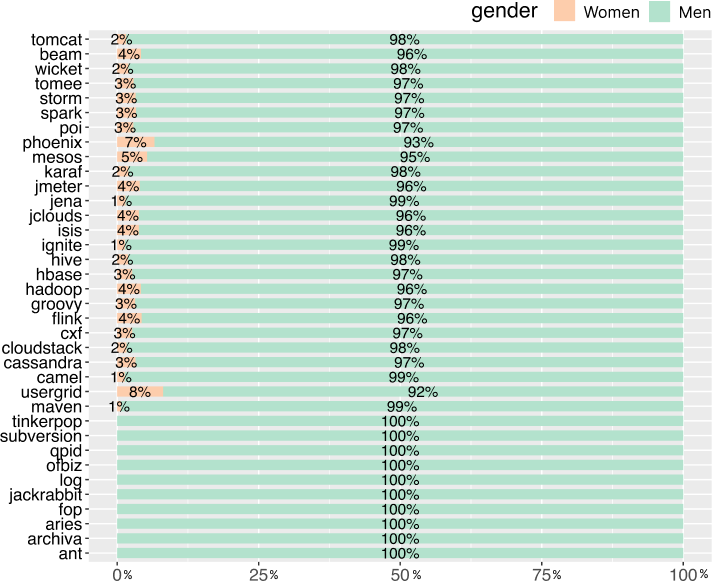}
\caption{Gender distributions of contributors in the projects in our dataset.}
\label{fig:gender_dist_plot}
\end{figure}

Past work has found \textit{homophily}---the tendency for people to seek out or be attracted to those who are similar to themselves---by gender among mentor-mentee pairs~\cite{ragins1999mentor}.

Before analyzing for homophily in implicit mentoring, we first analyzed the dataset to ensure whether there are women contributors in the projects for men to mentor. Figure~\ref{fig:gender_dist_plot} shows the gender distribution of contributors in the projects in our dataset. There were seven projects with no women contributors, and in one project ``Namsor" did not identify the gender of contributors with $>$90\% confidence, so we excluded these 10 projects from our analysis.

Table \ref{tab:gender_homophily_dist} (Overall column) presents the number of implicit mentoring PR-comments grouped by the genders of the mentor$\rightarrow$mentee pair. That is, w$\rightarrow$w means a woman mentored another woman and a w$\rightarrow$m means a woman mentored a man.  Two-sample proportional test based on differences of contributors' initial contribution date and differences in amount of work between homophilic implicit mentoring and cross-gender implicit mentoring show that these differences are significant ($P$ value$<$0.001) and the effect size is large (Cohen $d=2.55$). Similarly, homophilic mentoring was significantly more ($P$ value$<$0.001) than cross-gender mentoring for top-down and non-traditional (peer-to-peer, bottom-up) mentoring with large effect sizes for both mentoring approaches (Cohen $d>2.0$).

Recall, only seven of 231 survey respondents (3.03\%) identified themselves as female, which is nearly consistent with the results of the mining data analysis (3.17\%). We note that all of these women respondents agreed with the definition of implicit mentoring. All of women respondents who have been mentors and mentees were satisfied with their experiences. As one woman respondents said, ``[implicit mentoring] is the best way to train software engineers'' [ASF-126].

\begin{table}[tbp]
\caption{Implicit mentoring gender information.}
\resizebox{\columnwidth}{!}{
\begin{tabular}{lllll}
\rowcolor[HTML]{FFFFFF} 
\textbf{Gender} & \textbf{Overall} & \textbf{Top$\rightarrow$Down} & \multicolumn{2}{l}{\cellcolor[HTML]{FFFFFF}\textbf{Non-Top$\rightarrow$Down}} \\ \hline
\multicolumn{5}{c}{\textbf{Calcuated based on first contribution date}} \\ \hline
\rowcolor[HTML]{EFEFEF} 
\textbf{W$\rightarrow$W} & \multicolumn{1}{l|}{\cellcolor[HTML]{EFEFEF}3} & 1 & \multicolumn{2}{l}{\cellcolor[HTML]{EFEFEF}2} \\
\textbf{W$\rightarrow$M} & \multicolumn{1}{l|}{127} & 45  & \multicolumn{2}{l}{82} \\
\rowcolor[HTML]{EFEFEF} 
\textbf{M$\rightarrow$W} & \multicolumn{1}{l|}{\cellcolor[HTML]{EFEFEF}514} & 359  & \multicolumn{2}{l}{\cellcolor[HTML]{EFEFEF}155} \\
\textbf{M$\rightarrow$M} & \multicolumn{1}{l|}{25,103} & 13,438 & \multicolumn{2}{l}{11,665} \\
\rowcolor[HTML]{EFEFEF} 
\textbf{Total} & \multicolumn{1}{l|}{\cellcolor[HTML]{EFEFEF}25,747} & 13,843 & \multicolumn{2}{l}{\cellcolor[HTML]{EFEFEF}11,904} \\ \hline
\multicolumn{5}{c}{\textbf{Calcuated based on amount of contribution}} \\ \hline
\rowcolor[HTML]{EFEFEF} 
\textbf{W$\rightarrow$W} & \multicolumn{1}{l|}{\cellcolor[HTML]{EFEFEF}-} & 2 & \multicolumn{2}{l}{\cellcolor[HTML]{EFEFEF}1} \\
\textbf{W$\rightarrow$M} & \multicolumn{1}{l|}{-} & 59 & \multicolumn{2}{l}{68} \\
\rowcolor[HTML]{EFEFEF} 
\textbf{M$\rightarrow$W} & \multicolumn{1}{l|}{\cellcolor[HTML]{EFEFEF}-} & 433 & \multicolumn{2}{l}{\cellcolor[HTML]{EFEFEF}81} \\
\textbf{M$\rightarrow$M} & \multicolumn{1}{l|}{-} & 17,225 & \multicolumn{2}{l}{7,878} \\
\rowcolor[HTML]{EFEFEF} 
\textbf{Total} & \multicolumn{1}{l|}{\cellcolor[HTML]{EFEFEF}-} & 17,719 & \multicolumn{2}{l}{\cellcolor[HTML]{EFEFEF}8,028}
\end{tabular}}
\label{tab:gender_homophily_dist}
\end{table}

\begin{table}[!t]
\caption{Proportion test results of cross-gender mentoring}
\resizebox{\columnwidth}{!}{
\begin{tabular}{lllll}
\rowcolor[HTML]{FFFFFF} 
\textbf{Gender} & \textbf{Overall} & \textbf{Top$\rightarrow$Down} & \multicolumn{2}{l}{\cellcolor[HTML]{FFFFFF}\textbf{Non-Top$\rightarrow$Down}} \\ \hline
\multicolumn{5}{c}{\textbf{Calcuated based on first contribution date}} \\ \hline
\rowcolor[HTML]{EFEFEF} 
\textbf{Z-score} & \multicolumn{1}{l|}{\cellcolor[HTML]{EFEFEF}69.84} & 38.31 & \multicolumn{2}{l}{\cellcolor[HTML]{EFEFEF}62.97} \\
\textbf{P-value} & \multicolumn{1}{l|}{$<$ 0.01} & $<$ 0.01 & \multicolumn{2}{l}{$<$ 0.01} \\
\rowcolor[HTML]{EFEFEF} 
\textbf{\begin{tabular}[c]{@{}l@{}}Estimated\\ differences\end{tabular}} & \multicolumn{1}{l|}{\cellcolor[HTML]{EFEFEF}0.96} & 0.95 & \multicolumn{2}{l}{\cellcolor[HTML]{EFEFEF}0.96} \\
\textbf{Cohen d} & \multicolumn{1}{l|}{2.55} & 2.52 & \multicolumn{2}{l}{2.60} \\ \hline
\multicolumn{5}{c}{\textbf{Calcuated based on amount of contribution}} \\ \hline
\rowcolor[HTML]{EFEFEF} 
\textbf{Z-score} & \multicolumn{1}{l|}{\cellcolor[HTML]{EFEFEF}-} & 44.73 & \multicolumn{2}{l}{\cellcolor[HTML]{EFEFEF}59.77} \\
\textbf{P-value} & \multicolumn{1}{l|}{-} & $<$ 0.01 & \multicolumn{2}{l}{$<$ 0.01} \\
\rowcolor[HTML]{EFEFEF} 
\textbf{\begin{tabular}[c]{@{}l@{}}Estimated\\ differences\end{tabular}} & \multicolumn{1}{l|}{\cellcolor[HTML]{EFEFEF}-} & 0.94 & \multicolumn{2}{l}{\cellcolor[HTML]{EFEFEF}0.98} \\
\textbf{Cohen d} & \multicolumn{1}{l|}{-} & 2.46 & \multicolumn{2}{l}{2.70} \\
\rowcolor[HTML]{EFEFEF} 
\multicolumn{5}{l}{\cellcolor[HTML]{EFEFEF}$H_0: P_1-P_2 = 0$, $H_a: P_1-P_2 \neq 0$, adjusted $\alpha=0.017$} \\
\multicolumn{5}{l}{$P_1$: (m$\rightarrow$w) / (m$\rightarrow$m+ m$\rightarrow$w)} \\
\rowcolor[HTML]{EFEFEF} 
\multicolumn{5}{l}{\cellcolor[HTML]{EFEFEF}$P_2$: (w$\rightarrow$m) / (w$\rightarrow$w+ w$\rightarrow$m)}
\end{tabular}}
\label{tab:sec4_homophil_cross_gender}
\end{table}

\vspace{0.3cm}
\begin{mdframed}[roundcorner=10pt]
\textbf{Observation 5}:
There is a strong homophily effect in implicit mentoring, as there is still a significant disparity between the number of men and women contributors. 
\end{mdframed}


\section{Discussion}

Mentoring has been shown to be effective in helping with onboarding newcomers to OSS~\cite{fagerholm2014onboarding} as well as for existing contributors, as one of the interviewees, P3, said: ``\textit{I've gotten mentors, technically, my entire career in tech. And I mostly have looked for these mentors}''. In our recent study, we found that implicit mentoring occurs frequently in the daily activities of OSS contributors and is beneficial for both mentees and mentors \cite{feng2022}. To the best of our knowledge, we are the first to identify and investigate implicit mentoring in OSS.

\subsection{Non-top-to-down mentoring is the majority approach of implicit mentoring in OSS}
On looking deeper into who are the mentors and mentoring approach, non-top-to-down mentoring is the majority mentoring approach of implicit mentoring. P4 states that "... seniority is not necessary for implicit mentoring." These findings surprised us and remind us to keep investigating non-top-to-down mentoring. 
Traditional formal mentoring is the top to down approaching, but it is studied to indicate that formal mentoring programs are designed for newcomers \cite{fagerholm2014role,balali2018newcomers, balali2020recommending}. Our findings indicate that developers bring their experiences outsides of OSS and contribute to the OSS community, and help each other by sharing knowledge, giving suggestions/instructions, and identifying errors. The online communities can also leverage the classifier to identify and acknowledge implicit mentoring.

\subsection{Homophily in Implicit Mentoring}
In our study, we find the occurrence of homophily in implicit mentor-mentee pairs especially for men. There might be multiple reasons behind this, such as unconscious bias against other genders, personal preferences or mentor and mentee's common expectations from the relationship. As McPherson et al.~\cite{mcpherson2001birds} found, connections and friendships are based on social processes and personal preferences and are not randomly made. Therefore, in non-random mentorships, it is more likely to find homophily than it is when assigned formally. Another reason could be the mentee's view of same-gender mentors as being able to empathize with issues specific to their gender~\cite{RAGINS1990321}. The extensive amount of homophily in (implicit) mentoring ($>90\%$) amplifies an important call to action of improving diversity in OSS. The already low number of women in OSS and the deleterious effects of cross-gender mentoring on women creates a negative feedback loop that further disadvantages women in OSS.

\subsection{Encouraging Participation in Implicit Mentoring}

Volunteers in OSS projects contribute to the development of the software they use and the enhancement of their skill sets and resumes \cite{sarma2016training}. In this study, we found that implicit mentoring is not dyadic or top-to-down. Since, by its very nature, implicit mentoring constitutes a brief interaction, it is possible that if OSS projects recognize this mentoring, mentoring can become more sustainable. It is also possible that the brief interactions characterized by implicit mentoring can be transformed into longer-lasting interactions. Research has shown that informal mentoring relationships are more satisfying and often rooted in friendship~\cite{baugh2007formal, inzer2005review,bynum2015power}. One of the problems with mentor-mentee matching is diverting interests. Given that implicit mentoring is prevalent in OSS, it is perhaps feasible to match mentor-mentees together who share a common passion for some technical aspects of the project. Therefore, an OSS project might leverage our approach to facilitate implicit mentorship by adding a topic or needed skill tag, such that after submitting a PR, a contributor could create a topic tag coupled with (e.g. \#Machine Learning modelling) to attract the attention of contributors with similar interests.

A recurring problem raised by our interviewees and literature is mentor ``burnout'' and work overload~\cite{balali2018newcomers, fagerholm2014onboarding, trinkenreich2020hidden}. 
In our recent work \cite{feng2022}, we found as a result of supporting others, implicit mentoring enables mentors to continuously enhance their technical and their non-technical network. Mullen and Klimaitis~\cite{mullen2021defining} identified effective alternate mentoring models, such as group mentoring or collaborative mentoring. Such a collaborative mentoring model can reduce the workload among mentors, while at the same time creating a cohort of like-minded individuals who support each other. However, P4 said: ``\textit{there is no recognition, no kudos, no kind of positive reinforcement for them to continue being a mentor. So, usually, they are a mentor once and then they leave}''. Therefore, in order to increase mentor engagement, the OSS community should consider recognizing and rewarding mentors, such as using karma points award. As one survey respondent explained their motivation for implicit mentoring: ``\textit{Getting karma in the project. Typically becoming a committer or PMC member}'' [ASF-83].

Previous research has discovered that the percentage of negative sentiments in code review comments is higher than the ratio of positive sentiments and that destructive criticism and negative feedback are common \cite{paul2019expressions, gunawardena2022destructive}. Steinmacher et.al \cite{steinmacher2015social} found mentoring social barriers that slowed mentees' contributions, which may also affect mentoring programs for mentors, such as making irrelevant comments on the mailing list and not acknowledging answers. Therefore, communication issues between mentees and mentors might harm implicit mentoring engagement. Creating a code of conduct can explicitly state communication rules for both mentors and mentees, such as thanking mentors for their efforts, etc \cite{steinmacher2015social}.

OSS researchers can also investigate implicit mentors in more depth. For example, what motivates mentors? Past work has found that motivations to join OSS versus remain in OSS changes. As contributors become experienced members, their motivations change from extrinsic to intrinsic~\cite{gerosa2021motivation, feng2022}. Another question is to what extent do the professional and topical interests of the mentors and mentees need to align? For example, P2 stated ``\textit{...why would I help them unless they are going to help me achieve my own personal my own business goals?}'' Finally, researchers can explore what barriers exist for implicit mentors for giving feedback within the constraints of code review tools.

\section{Threats to Validity}
\label{sec:threats}

Like any other empirical research, our study also has threats. We have taken all possible measures to offset the impact of these potential threats as we detail below.

\textit{\textbf{Construct validity:}}
It is possible that some of the analyzed projects may have used other communication channels such as JIRA that we did not detect from the PR analysis. However, respondents to the survey confirmed that PR is the most popular channel they engaged in implicit mentoring.

There is always a threat to the construct validity if participants misunderstand our survey questions. To mitigate this threat, we conducted pilot studies with developers with different experience levels from OSS community. We updated the survey based on the feedback of these pilot studies.

One threat to validity can occur due to our use of the ``Namsor" API~\cite{namsor} to identify the gender of mentors and mentees in our dataset. ``Namsor" requires a person's full name and geographic location to predict a gender. However, not all GitHub contributors provide this information due to privacy concerns, which might cause noise in our data. To mitigate this threat, we first used ``Namsor'' to predict the origin of the contributor based on their name. Then, using both the predicted origin and the contributor's name, we predicted their gender, and only extracted the result if the prediction confidence is higher than 90\%, as previously done in related research \cite{santamaria2018comparison}.

\textit{\textbf{Internal validity:}} The manual analysis applied while labeling PR-comments could have introduced unintentional bias. To address this concern, two researchers individually labeled a significant portion of the data. We established a high Inter-Rater Reliability of 90\%, which, according to \cite{McHugh2012InterraterRT}, is considered as a substantial level of agreement. We then took samples from the manually categorized data and confirmed our understanding using member-checking as well as large scale surveys.

\textit{\textbf{External validity:}} We only covered PRs on Github Apache projects, but we obtained large samples. The dataset used for the study consisted of only Apache projects on GitHub. It is possible that the conclusions from our analysis may not apply to other OSS projects.  Similarly, we surveyed developers only from the Apache Software Foundation. The characteristics of these developers may not be representative of developers in other OSS projects.


\section{Conclusions and Future Work}


In this journal extension, we show that implicit mentoring bucks the traditional dyadic, top-down mentoring model, comprising a large portion of non-traditional mentoring. Given the large amount of implicit mentoring taking place that is currently unacknowledged in a project, mechanisms to acknowledge implicit mentors through badging or other mechanisms of appreciation can help make mentoring sustainable. Our results also show that, similar to other mentorship models, homophily was dominant in implicit mentoring ($>90\%$ of all implicit mentoring), especially for men. While this is expected, it raises serious concerns.

In our future work, we plan to address how implicit mentoring impacts organizations and projects, such as project health. We will also investigate the effects of implicit mentoring on diversity in OSS and whether fostering lightweight mentor networks can help to reduce the long-standing problem of lack of women's participation in OSS.

The research artifacts for this study are available publicly at the companion website~\cite{supply}


%

\appendices

\bibliographystyle{IEEEtran}
\bibliography{bib}

%

\begin{IEEEbiography}
[{\includegraphics[width=0.8in,height=1in,clip,keepaspectratio]{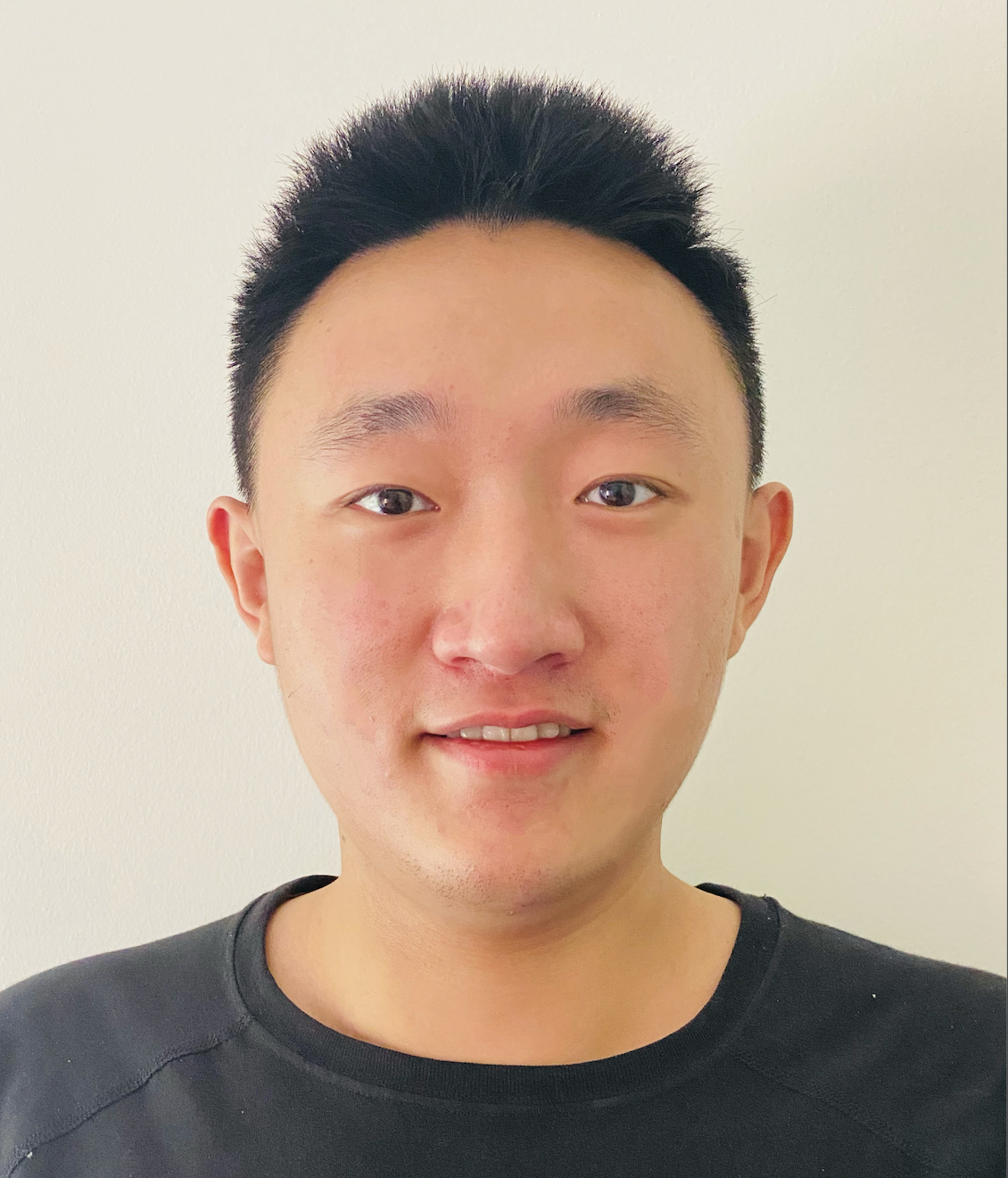}}]{Zixuan Feng} is a Ph.D. student at Oregon State University in the School of Electrical Engineering and Computer science. His research interest is in supporting collaboration in distributed teams on software engineering and data mining.
\end{IEEEbiography}

\begin{IEEEbiography}
[{\includegraphics[width=0.8in,height=1.5in,clip,keepaspectratio]{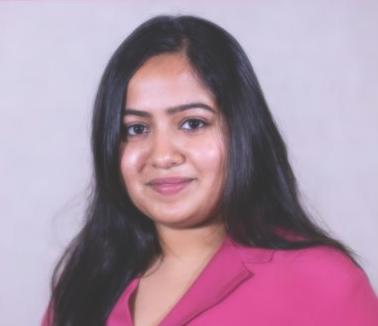}}]{Amreeta chatterjee} is Ph.D student in the School of Electrical Engineering and Computer science, at Oregon State University. Her research lies in the intersection of Software Engineering (SE) and Human Computer Interaction (HCI) and her interests include making software more inclusive to diverse users.
\end{IEEEbiography}

\begin{IEEEbiography}
[{\includegraphics[width=0.8in,height=1in,clip,keepaspectratio]{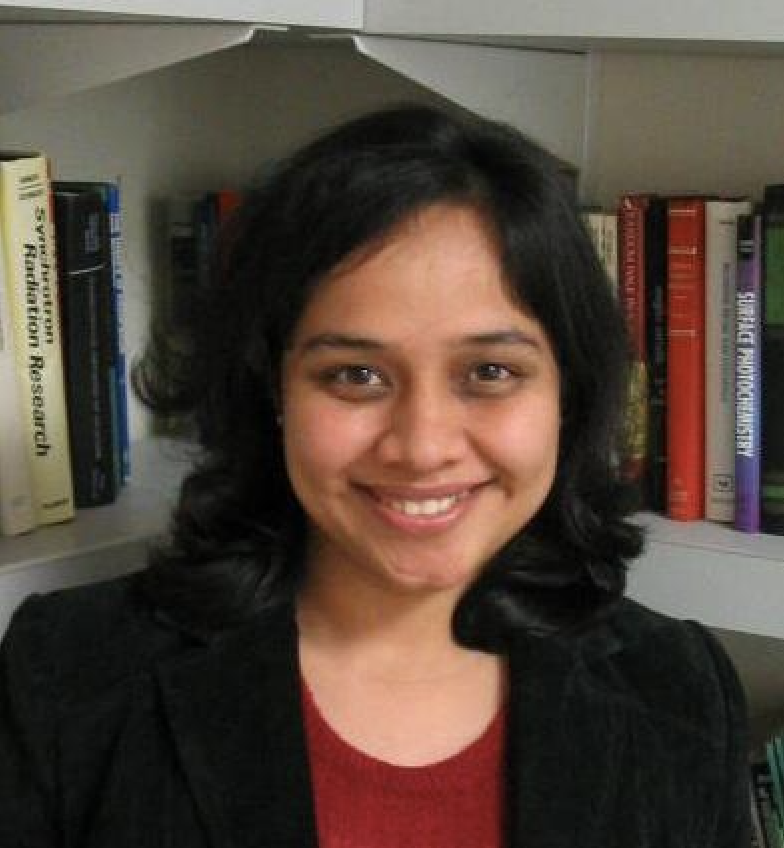}}]{Anita Sarma} is an Associate Professor at Oregon State University in the School of Electrical Engineering and Computer science. Her research investigate human factors in software engineering. She has over 100 papers in journals and conferences. Her work has been recognized by an NSF CAREER award as well as several best paper awards.
\end{IEEEbiography}

\begin{IEEEbiography}
[{\includegraphics[width=0.8 in,height=1in,clip,keepaspectratio]{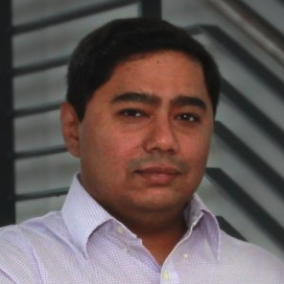}}]{Iftekhar Ahmed} is an Assistant Professor of Informatics in the
Donald Bren School of Information and Computer Science at the
University of California, Irvine. His research focuses on software
engineering in general and combining software testing, analysis and
data mining to come up with better tools and techniques in particular. He finished his B.Sc. in Computer Science and Engineering
from Shahjalal University of Science and Technology, Bangladesh
and after working in the industry for 4 years, did his Ph.D. at Oregon
State University.

\end{IEEEbiography}







\end{document}